\documentclass[aps,preprint,amsmath,amssymb]{revtex4}
\usepackage{psfig}% Include figure files
\usepackage{graphicx}

\begin{document}

\title{\large \bf
Impact of CP phases on a light sbottom and gluino sector}
\date{\today}

\author{ \bf Chuan-Hung Chen}
%$^{a}$ and C.~Q.~Geng$^{b,c}$ }

\affiliation{
% $^{a}$
Department of Physics, National Cheng-Kung University, Tainan 701,
Taiwan
%$^{b}$Department of Physics, National Tsing Hua
%University, Hsinchu, Taiwan 300, Republic of China \\
%$^{c}$ Theory Group, TRIUMF, 4004 Wesbrook Mall, Vancouver, B.C.
%V6T 2A3, Canada }
}

\begin{abstract}
We study a scenario in which light bottom squarks and light
gluinos with masses in the range $2-5.5$ GeV and $12-16$ GeV,
respectively, can coexist in the MSSM, without being in conflict
with flavor-conserving low-energy observables. We find that in
such a scenario, the anomalous magnetic moment of a muon could be
as large as $10^{-9}$, if the theory conserves CP. However, if the
theory violates CP, we conclude that not both, the gluino and
bottom squark, can be light at the same time, after the neutron
electric dipole moment constraint on Weinberg's 3-gluon operator
has been taken into account.
\end{abstract}

%\begin{abstract}
%We study whether the light sbottom and gluino with mass ranges of
%$2.0\sim 5.5$ GeV and $12\sim 16$ GeV, respectively, can coexist
%on flavor unchanged low energy physics in CP conserving and
%violating cases. We find that the anomalous magnetic moment of
%muon could be ${\cal O }(10^{-9})$ if no CP violating phase is
%involved. However, with CP phases, only one of them could be light
%if we consider the Weinberg's operator for the electric dipole
%moment of neutron.
%\end{abstract}

\maketitle

Inspired by the unforeseen excess of the bottom-quark production
observed at the hadronic collider of Fermilab \cite{Fermilab}, Berger
{\it et al.} \cite{Berger-PRL} proposed a solution with  light
sbottom $\tilde{b}_{1}$ ($m_{\tilde{b}_1}\simeq 2.0- 5.5$ GeV) and light
gluino $\tilde{g}$ ($m_{\tilde{g}}\simeq 12-16$ GeV).
% to try to understand the discrepancy.
Interestingly,
%the
%consequent
various problems
% on accurate experiments
can easily be avoided, such as by adopting the proper mixing angle
of two sbottoms $\tilde{b}_{L}$ and $\tilde{b}_R$, the Z-peak
constraint can be evaded; and also, the $R_{b}$ contribution,
arisen from sbottom-gluino loop, can be suppressed by considering
the second sbottom $\tilde{b}_{2}$ being lighter than $180$ GeV
for the CP conserved case \cite{CXY}, while the mass of
$\tilde{b}_{2}$ can be heavier in the CP violating one
\cite{Baek}. The amazing thing is that such light supersymmetric
particles, so far, haven't been excluded by experiments, even
after including the data of precise measurements \cite{DD}.
Moreover, for searching the signals of the light sbottom and
gluino, many testable proposals are raised, e.g., the rate of
$\chi_{b}$ decaying to a pair of light bottom squarks
\cite{Berger-PRD}, radiative $B$ meson decays \cite{Becher-PLB},
the decays $Z\to b \tilde{b}_{1}^{*} \tilde{g} +\bar{b}
\tilde{b}_{1} \tilde{g}$ followed by $\tilde{g} \to b
\tilde{b}_{1}^{*}/\bar{b} \tilde{b}_{1}$ and $e^{+} e^{-} \to
q\bar{q} \tilde{g} \tilde{g}$ \cite{CK}, as well as the running of
strong coupling constant $\alpha_{s}$ \cite{Chiang-PRD}.

To further explore more impacts on other processes,
%% whether the possibility has more impacts on the
%other processes or could be excluded by other physical phenomena,
it is necessary to investigate different systems instead of those
in which the final states are directly associated with the light
sbottom and  gluino.  It is known that supersymmetric models not
only supply an elegant mechanism for the electroweak symmetry
breaking and a solution to the hierarchy problem, but also
guarantee the unification of gauge couplings at the scale of GUTs
\cite{SUSY-GUTS}. Therefore, besides the effects mentioned above,
other contributions will also appear when considering different
phenomena. Inevitably, new parameters will come out. To avoid
introducing the irrelevant parameters, such as the mixing angles
among different flavors of squark, we have to consider the
processes in which the dependent parameters are still concentrated
on the minimal set. The best candidates are the flavor-conserving
processes.

One of the mysteries in the standard model (SM) is whether the
Higgs mechanism plays an essential role for the  symmetry breaking
and the resultant of Higgs particle can be captured in future
colliders. Based on the same philosophy of the symmetry breaking,
the minimal supersymmetric standard model (MSSM) needs the second
Higgs doublet field to balance the anomaly of quantum corrections,
i.e., there are three neutral Higgs particles in MSSM, one of them
is CP-odd ($A^{0}$) and the remains are CP-even ($h \text{ and } H
$). It is obvious that besides the mixing angle of sbottoms and
the masses of sbottom (gluino), the essential parameters in MSSM
are $A_{t(b)}$, the trilinear SUSY soft breaking terms, $\mu$, the
mixing parameter of two Higgs superfields, $m_{A^{0}(h,H)}$, the
masses of corresponding Higgses, and $\tan\beta$, defined by the
ratio of $v_{u}$ to $v_{d}$ in with $v_{u(d)}$ being the vacuum
expectation of the Higgs field that couples to up (down) type
quarks. And also, unlike the non-SUSY two-Higgs-doublet model, the
mixing angle of two neutral Higgs fields, denoted by $\alpha$, is
not an independent parameter and can be related to $\tan\beta$
\cite{Higgs-Hunter}.

The activity of  searching for the Higgs particle and studying its
properties is proceeding continuously \cite{Carena,Gunion}. In
particular, the remarkable results with a large $\tan\beta$ have
been investigated enormously because the exclusive characteristic
can give the unification of bottom and tau Yukawa couplings and
the realization of the top to the bottom mass ratio in GUTs
\cite{GUTs}. It has been found that if the SUSY soft breaking
terms carry the explicit CP violating phases, due to the
enhancement of large $A_{t(b)}$ and $\mu$, radiative effects can
induce a sizable mixing between scalar and pseudoscalar such that
the lightest Higgs boson could be $60-70$ GeV and thus escape the
detection of detectors \cite{Pilaftsis-Higgs}. Moreover, with the
requirements satisfied with electroweak baryogenesis, a novel
prediction on the muon electric dipole moment (EDM) of $10^{-24}\
e\; cm$ can be reached by the proposed experiment
\cite{Pilaftsis-NPB}. In sum, it will be more exciting that if the
light sbottom and gluino can be compatible with the Higgs physics
with or without CP violation (CPV).

In order to further pursue the implications of the light sbottom
and gluino, in this paper, we concentrate on two flavor-conserving
processes: one is anomalous magnetic moment of muon, $\Delta
a_{\mu}$, and the other  is EDMs of muon and neutron.
% in which
The former corresponds to CP conservation (CPC) while the latter
is related to CPV. Since the results must be proportional to the
mixing of $\tilde{b}_{L}$ and $\tilde{b}_{R}$, for generality, we
describe the relationship between weak  and physical
eigenstates as %%
\begin{eqnarray}
\left(\begin{array}{c}
  \tilde{b}_{L} \\
  \tilde{b}_{R} \\
\end{array} \right) =
\left(\begin{array}{cc}
  1 & 0 \\
  0 & e^{i\delta_{b}} \\
\end{array}\right)%%
\left( \begin{array}{cc}
  \cos\theta_{b} & \sin\theta_{b} \\
  -\sin\theta_{b} & \cos\theta_{b} \\
\end{array} \right)%%
\left(\begin{array}{c}
  \tilde{b}_{1} \\
  \tilde{b}_{2} \\
\end{array} \right),
\end{eqnarray}
where $\delta_{b}$ is the CP violating phase which could arise from
the off-diagonal mass matrix element $m_{b}(A_{b}^{*}-\mu
\tan\beta)$. To suppress the coupling $Z\tilde{b}_{1}
\tilde{b}_{1}$, we set $\sin2\theta_{b}=0.76$ in our
% whole
discussions.
%To be more clear,
In the following analyses, we separate the problem into CPC and
CPV cases.

\noindent {\it 1.} CPC ($\delta_{b}=0$):

%Differ from
Unlike non-SUSY two-Higgs doublet models, it is well known that
besides the enhancement of large $\tan\beta$ or $1/\cos\beta$,
there exists another enhanced factor $A_{b}$ in the couplings of
$S=h$ and $H$ to bottom squarks \cite{Higgs-Hunter}. As described
early, some novel consequences based on both large factors have
been displayed on the Higgs hunting. With these factors, we
examine the implication on $\Delta a_{\mu}$ by considering the
case of a light sbottom. In addition, for convenience, we adopt
the decoupling limit with $m_{A^{0}} >> m_{Z}$ so that
$\tan2\alpha \approx \tan2\beta$.
% Hence, we set
%$\alpha \approx \beta$ in our consideration.

 %%%%%%%%%%%%%%Figure %%%%%%%%%%%%%%%%%
\begin{figure}[hbt]
\includegraphics*[width=1.5
  in]{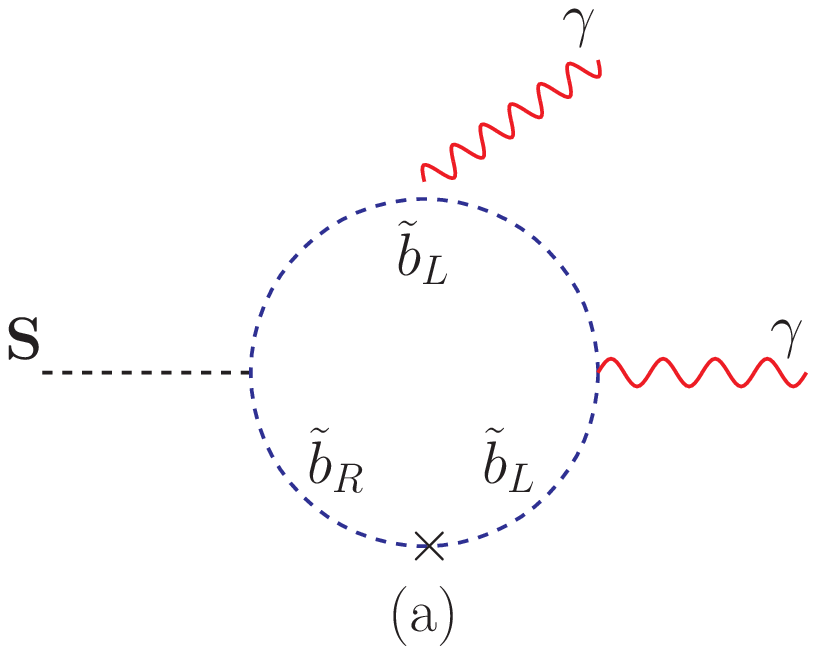} \includegraphics*[width=1.5
  in]{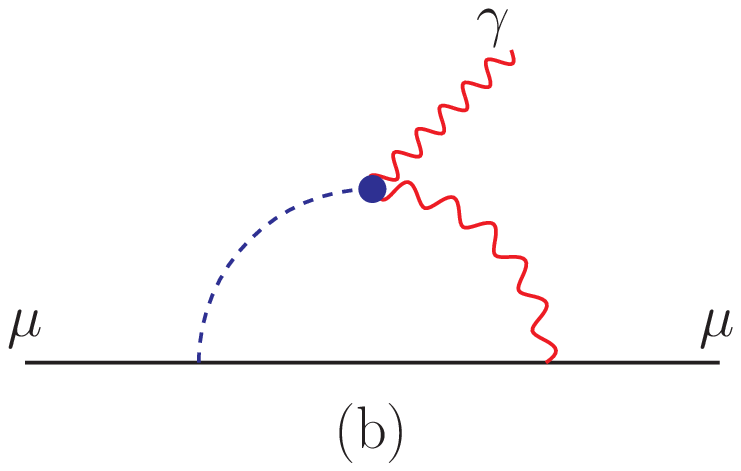} % Here is how to import EPS art
%\centerline{ \psfig{figure=brdf0.eps,height=1.6 in }
%\psfig{figure=brjf0.eps,height=1.6 in}}
\caption{ (a) Diagram which induces the effective coupling
$S-\gamma-\gamma$,
% induced by the first loop effects, in which
where  $S$ can be the lightest
scalar boson $h$ or pseudo-scalar boson $A^{0}$; (b) loop for
$\Delta a_{\mu}$. }\label{feyng-2}
\end{figure}
%%%%%%%%%%%%%%%%%%%%%%%%%%%%%%%%%%%%%%%%%%%%%%%

%First,
The effective interaction $h\gamma \gamma$ is induced from
the radiative effects in which sbottoms are the internal particles
of the loop, illustrated in Fig. \ref{feyng-2}(a), and the gauge
invariant form of the coupling can be obtained as
%\begin{eqnarray}
%{\cal L}&=&-\frac{ m_{b}}{v }\frac{(A_{b} \sin\alpha-\mu
%\cos\alpha)}{\cos\beta} \tilde{b}^{*}_{L} \tilde{b}_{R} h
%\nonumber \\
%&& -i \frac{ m_{b}}{v }(A_{b}\tan\beta-\mu)\tilde{b}^{*}_{L}
%\tilde{b}_{R} A^{0}+h.c.
%\end{eqnarray}
\begin{eqnarray}
i\Gamma^{\gamma}_{\mu\nu} &=& iA^{\gamma}(q^{2})[(q\cdot
k)g_{\mu\nu}-q_{\mu} k_{\nu}], \nonumber \\
A^{\gamma}(q^2)&=&N_{c}{\alpha_{em} Q^{2}_{b}m_b A_{b}\sin\alpha
\over 2\pi v\cos\beta}\sin2\theta_{b}
 \nonumber \\
&&\times \sum_{i=1,2} (-1)^{i+1} \int^{1}_{0} dx {x(1-x) \over
M^{2}_{\tilde{b}_{i}}-q^{2}x(1-x)}, \label{effc}
\end{eqnarray}
where $N_{c}=3$ is the color number, $\alpha_{em}$ is the fine
structure constant, $Q_{b}=-1/3$ is the charge of sbottom, and
$m_{b} (M_{\tilde{b}_{i}})$ is the mass of bottom quark (squarks).
Because we concentrate on the light sbottom case, we do not
discuss the contributions of stops by setting their masses being
heavy.
%and their effects could be
%neglected by setting heaviness.
Since $h$ couples to different sbottoms $\tilde{b}_{L}$ and
$\tilde{b}_{R}$ but $\gamma$ couples to the same sbottoms, it is
clearly inevitable to introduce the  mixing angle $\theta_{b}$.
 In Eq. (\ref{effc}), we already neglect the smaller
contribution related to $\mu \cos\alpha/\cos\beta$. By using the
result of Eq. (\ref{effc}), via the calculation of the loop in
Fig. \ref{feyng-2}(b), the anomalous magnetic moment of muon from
two-loop Higgs-sbottom-sbottom diagrams is given by
\cite{Chen-Geng}
\begin{eqnarray}
\Delta a_{\mu}&=&N_{c}\frac{
\alpha_{em}Q^2_{b}}{16\pi^3}\frac{m^{2}_{\mu}
m_{b}A_{b}\sin^{2}\alpha }{v^{2} M^2_{h}\cos^{2}\beta}
\sin2\theta_{b}\nonumber \\
&& \times  \sum_{i=1,2} (-1)^{i+1}F\left(
\frac{M^2_{\tilde{b}_{i}}}{M^{2}_{h}}\right),\nonumber \\
F(z)&=&\int^1_0 dx \frac{x(1-x)}{z-x(1-x)} \ln \frac{x(1-x)}{z}.
\label{eqg-2}
\end{eqnarray}
Immediately, we see that although this is a two-loop effect and
there appears one suppressed factor $m_{b}/v \cdot
(m_{u}/M_{h})^2$, the $\Delta a_{\mu}$ could be enhanced in terms
of $(A_{b}/v)\cdot \tan^{2}\beta$. Due to
$M_{h}>>M_{\tilde{b}_{1}}$, one finds that
$F(M^{2}_{\tilde{b}_{1}}/M^{2}_{h})\approx 2(1+\ln
M_{\tilde{b}_{1}}/M_{h})$. If we take $M_{\tilde{b}_{2}}=180$ GeV
and assume the lightest Higgs boson $100< M_{h}<140$ GeV, one gets
that $F(M^{2}_{\tilde{b}_{2}}/M^{2}_{h})\approx -0.25 \sim -0.16$.
Hence, it is certain that $F(M^{2}_{\tilde{b}_{1}}/M^{2}_{h})>>
F(M^{2}_{\tilde{b}_{2}}/M^{2}_{h})$.

\begin{figure}[hbt]
\includegraphics*[width=3.
  in]{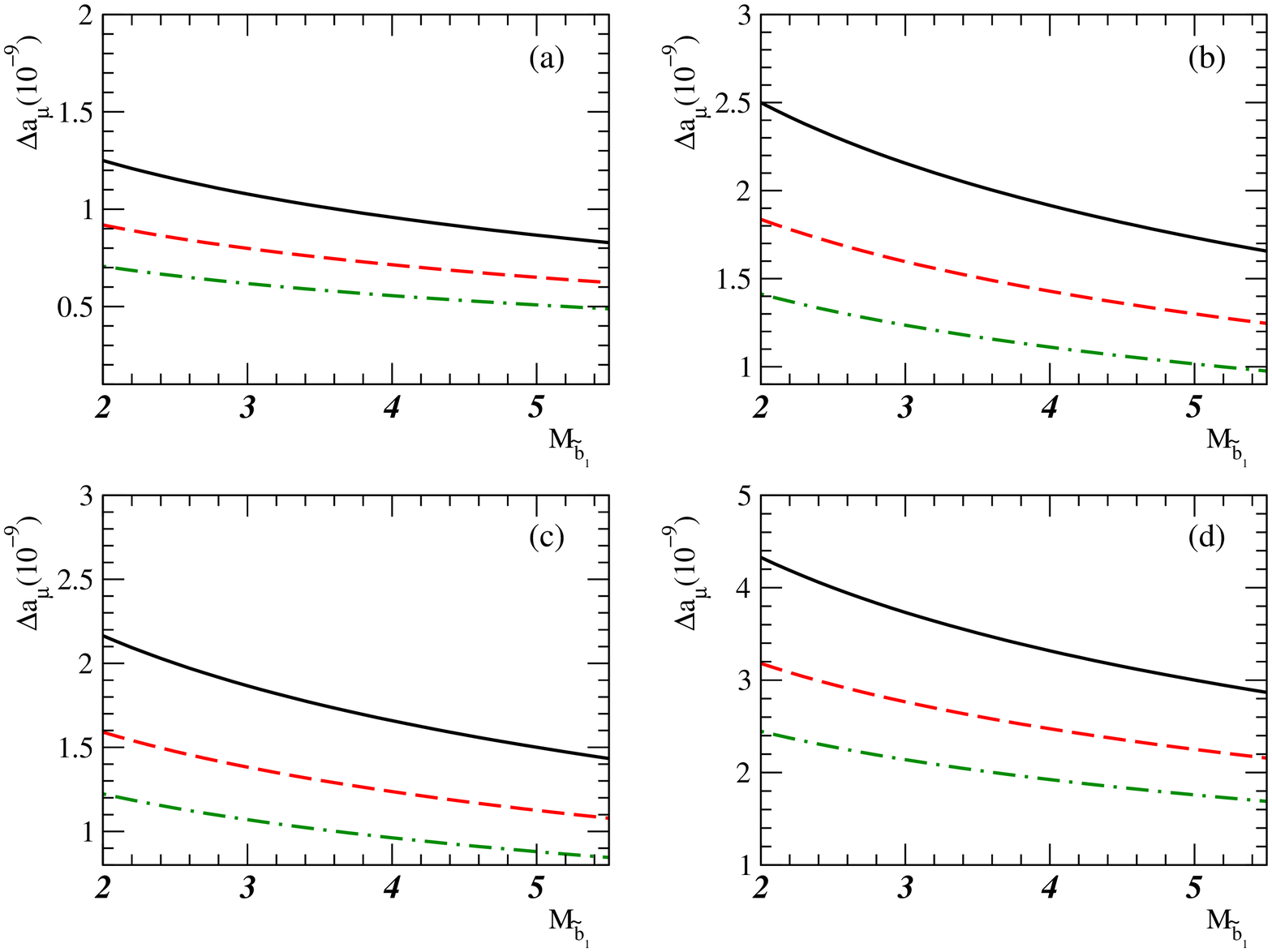} % Here is how to import EPS art
%\centerline{ \psfig{figure=brdf0.eps,height=1.6 in }
%\psfig{figure=brjf0.eps,height=1.6 in}}
\caption{ $\Delta a_{\mu}$ (in units of $10^{-9}$) for (a)[(c)]
$A_{b}=0.5$ TeV and (b)[(d)] $A_{b}=1$ TeV with
$\tan\beta=m_{t}/m_{b}$ [$\tan\beta=50$], where the solid,
dashed and dashed-dotted lines denote the results of $M_{h}=100,\
120$ and $140$ GeV, respectively.} \label{g-2}
\end{figure}

To understand the influence of free parameters
$M_{\tilde{b}_{1}(h)}$, $\tan\beta$ and $A_{b}$, we present the
$\Delta a_{\mu}$ as a function of $M_{\tilde{b}_{1}}$ for
$A_{b}=0.5$ and $1$ TeV with $\tan\beta=m_{t}/m_{b}$
[$\tan\beta=50$] in Fig. \ref{g-2}(a) and (b) [(c) and (d)], where
the solid, dashed and dashed-dotted lines denote the results of
$M_{h}=100,\ 120$ and $140$ GeV, respectively.
%As a result, we know that
Unlike non-SUSY models, which require the neutral Higgs to be a
few GeV to fit  $\Delta a_{\mu}$ \cite{Higgs-Hunter}, the lightest
Higgs boson in SUSY models could be the values constrained by the
current experiments. Interestingly, due to the light sbottom,  a
significant contribution to anomalous magnetic moment of muon
without extremely tuning $A_{b}$ is shown up \cite{Chen-Geng}.
 %%%%%%%%%%%%%%Figure %%%%%%%%%%%%%%%%%
%%%%%%%%%%%%%%%%%%%%%%%%%%%%%%%%%%%%%%%%%%%%%%%

\noindent {\it 2.} CPV($\delta_{b}\neq 0$):

CP problem has been investigated thoroughly in the kaon system
since it was discovered in 1964 \cite{CCFT}. Now, CPV has been
confirmed by Belle \cite{Belle-CP} and Babar \cite{Babar-CP} with
high accuracy in the $B$ system. Although the mechanism of
Kobayashi-Maskawa (KM) \cite{KM} phase in the SM is consistent
with the CP measurements, the requirement of the Higgs mass of
$60$ GeV for the condition of the matter-antimatter asymmetry has
been excluded by LEP. One of  candidates to deal with the
baryogenesis is to use the SUSY theory. As known, any CP violating
models will face the serious low energy constraints from EDMs of
lepton and neutron, which are $T$ and $P$ violating observables
and at elementary particle level usually are defined by
$d_{f}\bar{f} \sigma_{\mu \nu} \gamma_{5} f F^{\mu\nu}$, with
$F^{\mu\nu}$ being the electromagnetic (EM) field tensor. We note
that not only the EM field but also the chromoelectric dipole
moment (CEDM) of gluon for colored fermions will contribute.
Therefore, we have to examine the implication of the light sbottom
on EDMs while $\delta_{b}$ is nonzero, {\i.e.}, $A_{b}$ and $\mu$
are complex.
% in general.

\begin{figure}[hbt]
\includegraphics*[width=3.
in]{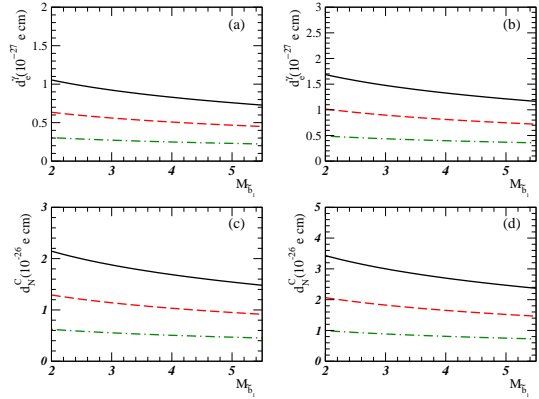} % Here is how to import EPS art
%\centerline{ \psfig{figure=brdf0.eps,height=1.6 in }
%\psfig{figure=brjf0.eps,height=1.6 in}}
\caption{ (a)[(b)] EDM of electron (in units of $10^{-27}$) and
(c)[(d)] CEDM of neutron (in units of $10^{-26}$) for
$\tan\beta=10$ and $A_{b}=500$ GeV [$\tan\beta=20$ and $A_{b}=200$
GeV] with $\delta_{b}=0.1$, where the solid, dashed and
dashed-dotted lines denote the results of $M_{A}=150,\ 200$ and
$250$ GeV, respectively. }\label{figedm}
\end{figure}

Inspired by the previous mechanism for $\Delta a_{\mu}$, the
similar effects with pseudoscalar $A^{0}$ instead of scalar $h$,
shown in Fig. \ref{feyng-2}(a), will also contribute the EDMs of
leptons and quarks. Since the effects have been analyzed by Refs.
\cite{Pilaftsis-PLB,Pilaftsis-NPB}, we directly summarize the
formalisms for the fermion EDM and CEDM  as
\begin{eqnarray}
\frac{d^{\gamma}_{f}}{e}&=&{N_{c}Q^2_{b}\alpha_{em} \over 64
\pi^3} {R_{f}Q_{f} m_{\ell} \over M^{2}_{A} }
\xi_{b}\sum_{i=1,2}(-1)^{i+1}
F\left(\frac{M^2_{\tilde{b}_{i}}}{M^{2}_{A}} \right),\nonumber \\
%%%%%%%%%%%%%%%%%%%%%%%%
\frac{d^{C}_{q}}{g_s}&=& \frac{1}{2Q_{b}^{2}Q_{f}N_{c}}
\frac{\alpha_{s}}{\alpha_{em}}\frac{m_{q}}{m_{f}}
\left(\frac{d^{\gamma}_{f}}{e}\right),
%%%%%%%%%%%%%%%%%%%%%%%%%
%\frac{d^{C}_{f}}{g_s}&=&{\alpha_s \over 128 \pi^3} {R_{f} m_f
%\over M^{2}_{a} }  \xi_{b}\sum_{i=1,2}(-1)^{i+1}
%F\left(\frac{M^2_{\tilde{b}_{i}}}{M^{2}_{a}} \right),
\label{EDM}
\end{eqnarray}
%%%%%%%%%%%%%%%%%%%%%%%%%%%%
respectively, with $R_{f}=\tan\beta\; (\cot\beta)$ for
$T^{f}_{3}=-1/2\;(1/2)$ and $\xi_{b}=\sin2\theta_{b} {m_{b}
\text{Im}(A_{b} e^{i\delta_{b}})}/( v^{2} \sin\beta \cos\beta)$.
For simplicity, we set the effects of stops to be negligible.
%\begin{eqnarray}
%\xi_{q}&=& {1 \over \cos^{2}\beta} {2m^{2}_{q} Im(\mu A_{q}) \over
%v^2 (M^2_{\tilde{q}_{2}}-M^2_{\tilde{q}_{1}})}.
%\end{eqnarray}
According to Eq. (\ref{EDM}), we see that the lepton EDM of
$d^{\gamma}_{\ell}$ is proportion to the $m_{\ell}$. Due to the
ratio $m_{\mu}/m_{e}\approx 205$, the strictest limit comes from
the EDM of electron, with the current upper bound being $1.6
\times 10^{-27}\ e\;cm$ \cite{eEDM}. Although the renormalization
factor $ (g_{s}(M_W)/g_{s}(\Lambda))^{32/23}$ for the EDM of
neutron is around one order larger than $
(g_{s}(m_{W})/g_{s}(\Lambda))^{74/23}$ for the CEDM contribution,
due to $\alpha_{s}$ and $1/(Q_{b}^{2}Q_{q})$ enhancements, the
CEDM of quark is much larger than the EDM of quark. Hence, only
the effects of the CEDM on neutron are considered. Since the
unknown parameters for the EDM of electron and CEDM of neutron are
the same, the values of parameters are taken to satisfy with the
bound of the electron EDM. We present the results as a function of
$M_{\tilde{b}_{1}}$ with $\tan\beta=10\; (20)$ and $A_{b}=500\;
(200)$ GeV in Fig. \ref{figedm}, in which the solid, dashed and
dashed-dotted lines correspond to $M_{h}=100,\ 120$ and $140$ GeV,
respectively. We note that the origin of CP violation comes from
$\text{Im}(A_{b}e^{i\delta_{b}})/|A_{b}|$ and it is set to be
${\cal {O}}(10^{-1})$. From the figure, we see clearly that
without fine tuning the CP phase to be tiny, the EDM of electron
can be lower than the  experimental bound and the CEDM of neutron
is also not too far away from current limit. The effects become
testable in experiments.

%%%%%%%%%%%%%%Figure %%%%%%%%%%%%%%%%%
%%%%%%%%%%%%%%%%%%%%%%%%%%%%%%%%%%%%%%%%%%%%%%%

It seems that so far the scenario of the light sbottom with the
mass of few GeV could give interesting results in $\Delta
a_{\mu}$, $d^{\gamma}_{e}$ and $d^{C}_{N}$. However, what we
question is whether both light sbottom and gluino can coexist when
the CP phases are involved. We find that the possibility
encounters fierce resistance of the Weinberg's 3-gluon operator,
defined by \cite{Weinberg}
\begin{eqnarray*}
{\cal O}=-\frac{1}{6} d^{G} f_{\alpha \beta \gamma} G_{\alpha \mu
\rho}G_{\beta \nu}{}^{\rho} G_{\gamma \lambda
\sigma}\varepsilon^{\mu\nu\lambda\sigma}
\end{eqnarray*}
with $G_{\alpha \mu \rho}$ and $f_{\alpha \beta \gamma}$ being the
gluon field-strength tensor and antisymmetric Gell-Mann
coefficient, respectively. With the bottom, sbottom and gluino
being the internal particles of the two-loop mechanism, the
contribution of the Weinberg's operator is given  by \cite{DDLP}
\begin{eqnarray}
d^{G}=-\frac{3\alpha_{s}m_{b}}{2}\left(
\frac{g_{s}}{4\pi}\right)^{3}
%Im(\Gamma^{12}_{b})
\sin2\theta_{b}
\frac{z_{1}-z_{2}}{M^3_{\tilde{g}}}H(z_{1},z_{2},z_{b})
\sin\delta_{b}
%+(b\rightarrowt)
\end{eqnarray}
where function $H$ is a two-loop integration,
$z_{i}=M^{2}_{\tilde{b}_{i}}/M^{2}_{\tilde{g}}$ and
$z_{b}=M^{2}_{b}/M^{2}_{\tilde{g}}$. Here, we have rotated away
the phase of gluino. With naive dimentional analysis, the neutron
EDM could be estimated by $d^{G}_{N}=(e\mu_{X}/4\pi)\eta_{G}
d^{G}$, with $\mu_{X}\sim 1.19$ GeV and $\eta_{G}\sim 3.4$  being
the chiral symmetry breaking scale and renormalization factor of
the Weinberg's operator, respectively. Although the estimation
still has a large theoretical uncertainty, our concern is on the
problem of order of magnitude.
% but not on a few factor.
Due to the result being proportional to $1/M^{3}_{\tilde{g}}$, we
see that the lighter $M_{\tilde{g}}$ is, the larger $d^{G}$. It is
worth mentioning that the similar tendency can be also found in
Ref. \cite {DPR}, in which the considered situation is via the
gluino$-$stop$-$top-quark two-loop. For an illustration, we
present the results in Table \ref{tableedm} with some given values
for $M_{\tilde{b}_1}$ and $M_{\tilde{g}}$ and fixing
$M_{\tilde{b}_{2}}=180$ GeV and $\tan\beta=10$.%%%%%%%%%%%%%%%%%%
\begin{table}[htb]
\caption{ Neutron EDM (in units of $10^{-25}\;e\;cm$) from
the Weinberg's operator with some taken values for $M_{\tilde{b}_1}$
and $M_{\tilde{g}}$ and fixing $M_{\tilde{b}_{2}}=180$ GeV and
$\tan\beta=10$. }\label{tableedm}
\begin{ruledtabular}
\begin{tabular}{ccccc}
 ($M_{\tilde{b}_{1}}$,$M_{\tilde{g}}$) & $(5, 16)$ &  $(5, 1000)$ & $(100,16)$ & $(100,1000)$ \\
\hline $d^{G}_{N}/\sin\delta_{b}$ &  $2.2\cdot 10^{4}$  & $8.2$  &
$1.7\cdot 10^{3}$ & $8.0$
\end{tabular}
\end{ruledtabular}
\end{table}
 From the table, we conclude  that unless the CP violating phase $\delta_{b}$
is tuned to be of
$O(10^{-4}-10^{-3})$, the scenario of light sbottom and
gluino with CPV will encounter the problem of naturalness.

In summary, we have extended the scenario of light sbottom to the
flavor-conserved low energy physics and shown that  by choosing
proper values of $\tan\beta$, $A_{b}$ and $M_{h(A)}$, the
predictions of $\Delta a_{\mu}$ and EDMs of electron and neutron
could satisfy with  experiments. We have also found that except
extreme fine tuning, the light sbottom and gluino cannot coexist
after the neutron electric dipole moment constraint on the
Weinberg's 3-gluon operator has been taken into account. We note
that it is possible that there exist some cancellations while we
consider all possible contributions to the neutron EDM \cite{BGK}.
Nevertheless, it will depend on how to fine tuning the involved
parameters and that is beyond our scope of the current paper.

\noindent {\bf Acknowledgments}

We thank C.Q. Geng, Otto C.W. Kong, C. Kao, K. Cheung and W.Y.
Keung for their useful discussions. This work was supported in
part by the National Science Council of the Republic of China
under
 Contract No. NSC-91-2112-M-001-053.
% and NSC-91-2112-M-007-043.

%%%%%%%%%%%%%%%%%%%%%%%%%%%%%%%%%%%%%%%%%%%%%%%%%%%%%%

\end{document}